# Detection of parallel steps in programs with arrays

R. Nuriyev
(renat.nuriyev@gmail.com)

The problem of detecting of information and logically independent (DILD) steps in programs is a key for equivalent program transformations. Here we are considering the problem of independence of loop iterations – the concentration of massive data processing and hence the most challenge construction for parallelizing. We introduced a separated form of loops when loop's body is a sequence of procedures each of them are used array's elements selected in a previous procedure. We prove that any loop may be algorithmically represented in this form and number of such procedures is invariant. We show that for this form of loop the steps connections are determined with some integer equations and hence the independence problem is algorithmically unsolvable if index expressions are more complex than cubical. . We suggest a modification of index semantics that made connection equations trivial and loops iterations can be executed in parallel.

We are considering not only algorithmic fullness of the programming language but also its data fullness when selection functions of elements of arrays are algorithmically full class too. These two features are independent. We suggest a modification of index semantics that made connection equations trivial and loops iterations can be executed in parallel. This modified language is a full in both senses.

We consider a DILD problem for programming languages with arrays and focusing on the syntax constructions determining the DILDs.

Transformations of programs, and parallelizing in particular, are based on fact that execution order for given start values can be changed without changing result. For program with single variables for fragment

    x=f(u)   // step 1
    y=g(x)   // step 2

we may say that step 2 informational depends on step 1.

But for the case of indexed variable
    x[i] =f(u)   // step 3
    y=g(x[j])   // step 4
the information dependency takes a place only if values i and j are the same.

Let's consider two program fragments, structured and unstructured in Dijkstra's terms, when each loop has one entrance and one exit points as in figure 1:

Figure 1

For the first fragment to identify an execution step for operator q we may use 2-dementional vector (p, k): p is a number of work out iterations of the upper loop and k is a number of lower loop iterations in the iteration p of upper loop. Step $q^{3,7}$ means that we are talking about such execution of the operator q when the lower loop made 7 iterations in a third iteration of the upper loop.

To identify a step for the second fragment we have to use variable length vector $q^{(i1,i2,...)}$ meaning that we are talking about point of execution when upper loop executes i1 times, then lower executes i2 times, then upper executes i3 times and then inner executes i4 times and so on.

Good news is that any program can be algorithmically structured (Dijkstra's **goto** elimination problem) so that only nested loops may be used.

In section 3 we will extend the canonization result to recursion programs.

So we will consider from now and up only **structured programs:** programs in which each repetition steps are formed only with nested loops.

We will extend this idea by introducing **forced syntax principle** approach to modify of the language that syntactically implicit properties we need to solve problem in hands are encapsulated in a (new if necessary) syntactically explicit constructions.

Applying this principle to the DILD problem with indexed variables we will introduce a representation of a loop body as a chain of special subprograms (last one is called **kernel**, others are called **controllers**) which are responsible for calculation values of indexes for variables of next levels controllers or kernel but not for themselves. Such organized loops are called **separated** forms of loops. Other word, in separated loops the two loop body's functions of selecting data from some sets or arrays and processing that data are separated. For programs with separated loops we have enough syntactical objects to be able to define a condition if some set of iterations can be run in parallel.

A separated loop for processing connected list has two level controllers. First level controller selects node and the second one selects pointer to the next element.

In this paper we'll prove that for FORTRAN like languages (or containing FORTRAN) the class of algorithmically parallelizing programs can't be remarkably bigger than the class with linear index expression. So to create language with algorithmically parallelizing programs we have to change some fundamental constructions of the languages and our goal is to point to these constructions.

**1. Basic definitions**

We will consider schemas of programs in which functions and predicates are symbols and they get interpretation (implementations) together with start data. But also we will allow some of them to have a fixed interpretation.

Let's A, X, F be finite alphabets without common elements and A*, X*, F* are a sets of strings in an appropriate alphabets.

**Def. 1. 1.** We call a **memory M** a countable set of **cells** - two types of elements: **x** and $a[w_1, .., w_m]$, where **x** belong to the set X* and is called single cell, *a* belongs to A* and is called array of cells, $w_1,…, w_n$ belong to X* and is called indexes of the cell.

A cell may get (be assigned to) a **value** from set $Y$ and keep it until next changing by a program operator. Cells without value are called **empty**. Each memory element has no more than one value and the value of element x will be denoted <x>, so <x>$\in Y$.

To keep this notation more close to programming languages let's suppose that one array name can't have different dimensions. Pairs of sets X, A and A, $Y$ do not have common elements.

**Def. 1.2. Variables** of a program schemas are elements of $X^*$ and expressions $a[K_1(z_1),\ldots, K_n(z_n)]$, where $a \in A^*$ is called array name, $K_1, \ldots, K_n \in F$ are called index functions, $z_1, \ldots, z_n$ are program variables from $X^*$. Variables of a first type are called **simple** and of a second type – **indexed variables**.

For X={x,y}, A={a,b}, expressions x, y, xx, xyx are a simple variables, expression ba[xyx, yx] is an example of indexed variables, expression aa[ba[x]] is not a variable – expression in [ ] parenthesis has to contain only simple variables.

**Def. 1.3. Schema of program** is a 5D vector **(Opers, F, VarX, VarA, subP)** where **VarX** is a set of single variables, **VarA** is a set of array variables, **F** is a set of function an predicate symbols, **subP** is a set of (sub) schemas, also called **procedures**, **Opers** is a finite set of program instructions – strings of one of the forms:
   a. $l_1$: $x_0$=g($x_1, \ldots, x_n$) **then** $l_2$; // assignment operator
   b. $l_1$: **if** p($x_1,\ldots,x_n$) **then** l2 **else** $l_3$;// conditional operator
   c. $l_1$: **do** $P_1$ **while** p($x_1,\ldots,x_n$) **then** $l_2$;// loop with body $P_1$ and iteration condition p($x_1,\ldots,x_n$),
   d. $l_1$: **do** $P_1$ **then** $l_2$; // call sub schema or procedure $P_1$.
Text after "//" is a comment, not a part of instructions.

Here $l_1, l_2, l_3$ are called **labels**, $l_1$ is called an **input label**, $l_2$ and $l_3$ are called **output labels**. $P_1$ is called a **sub schema** and set of it labels does not have common labels with upper level program or any other procedures, $p_1$ is called a **repetition predicate**, $g \in F$, $x_0, x_1, \ldots, x_n \in$ **VarX** $\cup$ **VarA**.

Output labels which are not input label are called **final,** and only one final label is allowed. Collection of labels of program or procedure P will be denote L(P).

One separate label $l_0$ is called **start label** and its operator – **start operator**.
We assume that each of sets **Opers, F, VarX, VarA, subP** includes such sets for procedures too.

Schema is called **determined** if its instructions have different input labels.

**Interpretation** of the schema is a map of some finite set of memory elements to elements from $Y$ (they called **start value** for the program), function symbols interpreted as maps [$Y^n \to Y$], predicate symbols interpreted as maps [$Y^n \to$ {true, false}].

Program variable **x** from **X** has a value <x> of the element of memory x, variable **a[K$_1$(x$_1$, …, x$_n$)…,K$_n$(x$_1$, …, x$_n$)]** has a value a[IK$_1$(<x$_1$>,.., <x$_n$>), …, IK$_n$(<x$_1$>,…, <x$_n$>)], IK$_j$ is an interpretation of K$_i$. Value of the variable is **empty** (not defined) if some of its index function has no value or one of memory element used as argument for index function is empty.

**The execution of operators** for a given interpretation consists of three parts - calculating some function or sub schema from the operator, changing the memory elements and marking some operator as next to execute.

More accurate, for operator of type:

**a**. calculate function Ig (interpretation of g) with <x$_1$>, …, <x$_n$> arguments and put the result to cell x$_0$, mark next to execute operator with input label **l$_2$**;

**b**. calculate Ip(<x$_1$>,…,<x$_n$>), if it is true - mark next to execute operator with label **l$_2$**, if it is false - mark next to execute operator with label **l$_3$**;

**c**. calculate sub schema P$_1$ and then if Ip(<x$_1$>, …, <x$_n$>) is true repeat this operator again, if it is false then mark next to execute operator with label **l$_2$**;

**d**. calculate sub schema P$_1$ and then mark next to execute operator with label **l$_2$**;

Operator execution is **defined** in following cases

**a.** all variables x$_1$, …, x$_n$ are defined and function Ig(<x$_1$>, …,<x$_n$>) is defined as well;

**b.** all variables x$_1$,…, x$_n$ are defined and predicate Ip(<x$_1$>,…,<x$_n$>) is also defined;

**c.** any iterations of P$_1$ are finished with final label and Ip(<x$_1$>,…,<x$_n$>) after each iteration has a value **true** and becomes **false** after finite number of steps;

**d.** sub schema P$_1$ stops in his final label.

**Execution of schema** for a given interpretation is a chain of executions of marked operators starting with operator with start label **l$_0$**.

Execution is considered **ended successfully** if after finite number of steps some of marked label is final label.

Execution **ended without result** if some supposed to be calculated predicate or function does not have value or one of its arguments does not have value.

**Remark 1.1**. It is possible that schema will **run forever** without reaching final label. So schemas calculate **partially defined function**.

For shortness we will omit letter I in interpretation of functions and predicates if it is clear from context what it has to be - symbol or its interpretation.

## 3. Loops in a separated form

Canonic forms of the studying objects always have a theoretical and practical value because they allow classify objects, to work with smaller diversity and to use smaller description of the objects. We saw it for step enumeration of structured program. Also it nicely comes with level of abstraction for program design.

We hope that studying a canonic form for information processes in programs will help to simplify design, develop better debugging tools and increase the code reusability.

In this section we will show that any loop body can be represent as a sequence of sub procedures determine values of array indexes for next procedures, but not for itself or upper level controllers. Last sub procedure called loop's **kernel**, others are called **controllers** of corresponding levels. So we separate the loop's body execution in two functions: hierarchical selecting data parts and parts processing them.

Such implementation allows to reduce a debugging of complex processes of data selection in arbitrary arrays to the chain of debugging a simple blocks without arrays. It comes from functions of controllers. If each upper level controller works properly, then data stream for this block works correctly and hence we need to check if this block is correct.

In C++ STL classes and generic collections in C# algorithms represent loops with fixed controllers for getting next elements from lists, maps, stacks and so on.

For theoretical studies this result is important because it gives syntactical objects for describing information connections in loops and, as it will be proved next, to divide the loops with different numbers of controllers or their connection graphs to unequal classes.

**Def. 3.1**. Let S(P) be a set of sub procedures of P, SS(P) be a transitive closure of this relations. Then sub procedure P called **recursive** if $P \in SS(P)$.

**Def .3.2.** Schema **P** is called **loop structured** (for short **L-schema**) if for **P** and for each of its sub schemas $P_i$:
  **a.** a relation "input label is less than output labels" is a partial order, called "structural",
  **b.** $P_i$ is not a recursive procedure.

So there is no recursion call, no spaghetti paths and no loops created by labels.

**Remark 3.1.** In L-schemas any repetition process can be generated only with nested loops.

**Remark 3.2.** L-schemas are structured in Dijkstra's terms: each loop has one start point and one exit point (we consider one final label only).

**Def. 3.3.** Two schemas with the same set of non interpreted symbols of functions and predicates are called **equal** if for each interpretation one schema finished successfully if and only if another schema finished successfully too and their state of memory is the same.

We can be proved the following
**Theorem 3.1.** There exists an algorithm to transform any schema to an equivalent L-schema.

Full proof is mostly technical, too bulky and obvious. Instructions that breach the label orders can be encapsulated in loops. Recursions can be replaced with pair loops and additional interpreted stack operations. Duplications of sub procedures can be eliminated by renaming and copying.

**Def. 3.4.** For a given instruction with input label **m** denote Ind(**m**), Arg(**m**), Val(**m**) sets of its index variables, set of its argument variables, set of its output variables. More accurately, for instruction **m** of following type

    a)    Ind(**m**) is a set of indexes of array variables $x_0, x_1, \ldots, x_n$; Arg(**m**) = $\{x_1,\ldots, x_n\} \cup$ Ind(**m**); Val(**m**)=$\{x_0\}$,

    b)    Ind(**m**) is a set of indexes of array variables from $\{x_1,\ldots,x_n\}$; Arg(**m**)=$\{x_1,\ldots,x_n\} \cup$ Ind(m);Val(m)=$\emptyset$,

    c)    Ind(**m**)=$\cup$Ind(k|k$\in$L(P)); Arg(m)=$\cup$Arg(k|k$\in$L(P))$\cup$Ind(**m**); Val(m)=$\cup$Val(k|k$\in$L(P));

    d)    Ind(**m**)=$\cup$Ind(k|k$\in$L(P)$\cup$J$_I$, where $J_1$ is set of indexes of array variables of predicate p; Arg(**m**)=$\cup$Arg(k|k$\in$L(P))$\cup$Ind(**m**); Val(**m**)=$\cup$Val(k|k$\in$L(P)),

For program or sub procedure P a set Ind(P)=$\cup$Ind(k|k$\in$P), Arg(P)=$\cup$Arg(k|k$\in$P)$\cup$Ind(P), Val(P)=$\cup$Val(k|k$\in$P).

**Def. 3.5.** Loop **C** with non empty set of index variables is called **separated** if its body P is a set of instructions

    $m_0$: do $P_1$ then $m_1$;

    …..

    $m_n$: do $P_2$ then $m_{n+1}$,

where $\forall i \forall j$ [(Ind($P_i$)$\cap$Val($P_{i+j}$)=$\emptyset$ ) &(Ind($P_i$)$\cap$Val($P_{i-1}$)$\neq \emptyset$)], i, j$\in$N.

Here $P_1,\ldots P_{k-1}$ are called **controllers** of levels $1,\ldots,k-1$, the last $P_k$ is called a **kernel** of the loop.

**Def. 3.6.** The separated loop called **strictly separated** if
$\forall P_i(i<k)$ Val($P_k$)$\cap$(Arg($P_i$)$\cup$Ind($P_i$))=$\emptyset$.

Other words, a kernel of a strictly separated loop does not change any variable used by any controller.

**Def. 3.7.** Schema is called **simple** if it consists only of instructions of type a) and d).

**Def. 3.8.** Two schemas called **t-equally** if they are equal for each interpretation where functions and predicates are define for each arguments (also called **totally define**).

**Def .3.9.** Schema is called **forward oriented** if each loop has a body in which index of variable for any interpretation can be changed only before using. Syntactically, for each branch with array variables there is no operator changing its indexes with bigger label.

The following auxiliary statement can be proven.

**Lemma 3.1.** There exists an algorithm of transformation of any L-schema to the t-equal forward oriented schema.

**Idea of proof.** Let's have two instructions $S_1$ and $S_2$ in one branch of P. $S_1$ is executed early (input label of $S_1$ is less than for $S_2$) and uses index variable E and $S_2$ is executed later and changes E in one of the branches. Then we'll add new variable newE and add ahead of $S_1$ instruction
   newE = E;
and replace index E to newE in $S_1$.

Modified branch looks like next
….
newE=E;
x1=f(x[.., newE,..]…);
….
E=g(…);
…
It clearly equals to old branch for any interpretation of f and g and the branch now satisfies for forward orientation condition. By repeating such modification for any branch and loop body we will end up with forward oriented schema.

Now we are ready for the main result of this section.

**Theorem 3.2.** There exists an algorithm to transform any loop C with arrays to t-equally separated loops with some **n** controllers and there is no any equally separated loop with different number of controllers.

**Proof of the first part of theorem**. Let's B be a body of the loop C. According to lemma 2.1 we may assume that B has a "structural" partial order "<" on L(B). Let's for each branch collect such of instructions k with bigger input label than start instruction in order "<" and built a set of left side variables Vs = $\cup$Val(k) until we meet instruction with indexes from Vs or get a final one. Let's call this instruction "limited", and continue with other branches. Process stops when all branches are visited and each ended with limited (red) or the final instruction. Visited set of instructions constitutes a first level controller.

To finish building the first controller let's add interpreted instructions with a new variable vLeb1 which will keep the output label of last executed instructions. It may looks like the following. Let $m_r$ is a final or limited output label. Then we'll replace $m_r$ with new label $m_{Aux}$ and add instruction
   $m_{Aux}$:  vLab1='$m_r$'.

We also add the next interpreted instructions to the rest set of instructions after removing organizer's instructions:

$m_{aux0}$: if (vLeb1==$m_1$) then $m_1$ else $m_{aux1}$ // start instruction of next procedure

$m_{aux1}$: if (vLeb1==$m_2$) then $m_2$ else $m_{aux2}$
$m_{aux2}$: ..,
here $m_{auxi}$ are new auxiliary labels.

These additions guarantee that after execution of the first controller the calculations will continue in an original order.

Clear, that if the rest set of instructions has instruction which changes indexes of others we may repeat the above process: mark as limited those instructions that have indexes from $\cup Ind(k|k \in L(P)) - Val(P_0)$ and separate next level controller $P_i$. If there are no such instructions then the loop has only i controllers and the set of rest instructions is a kernel.

Second part of the statement (about number of controllers) can be proven using special interpretations which we borrowed from Gödel's model theory. He developed it to study logical model (and called it model on constants). We will use this technique several times latter in this study.

Suppose we have some formal system with signature consisting of symbols of functions from **F** and predicates from **P**. Then the set $Y$ of values for standard interpretations is a set of terms T, built with elements of F and variable expressions. Formally T can be defined by induction:
1. Simple variables X, used in schema (finite set) are elements of T;
2. If $a[r(x_1,.., x_{n1}), ..]$ is an array variable in schema, then expression (string of symbols) '$a[r_1(x_1,…, x_{n1}), ..]$' is in T for any $x_1,…, x_n$ from T.
3. term $f(t1,…,tn)$ for each $t1,..tn$ from T and $f \in F$ also belongs to T.

**Standard interpretation of a functional symbol** f with n arguments is a map terms $t_1,…, t_n$ to the term (chain of symbols) '$f(t_1,…, t_n)$' if it is in T and is not define if it is not in T. In other word, the value of a function is a path of its calculation.

**Standard interpretation of a predicate symbol** is determined with finite set D, called a **diagram**, of expressions '$p(t_1,…,t_n)$' or '$\neg p(t_1,…,t_n)$' (only one of these two expressions is allowed to be in D ), where $t_1, …,t_n$ are from T.

Now we may prove the second part of the Theorem 2.2 that number of controllers is invariant for equivalent transformations.

Let's consider loop C for standard interpretation. By construction, each controller changes at least ones the indexes used in next organizer. So the value of indexes for standard interpretation have to be a word as '$a[…, t,…]$' where term t is a value of previous controller. Next level controller has to have a branch which changes indexes for next after it level and has to contain this expression. Otherwise if each path (for total schema it is also a branch) of such controller does not have this word in expression t for index, then all paths must be included in the previous levels.
Therefore execution of n controllers has to have value with n-1 deepness of parenthesis []. Hence two loops with different number of controllers can't be equal.

Proof is finished.

This technique may be used for more detailed classification of the loops. For example, from the proof it also follows that loops with different dependency graph between controllers also can't be equal.

## 4. Immediate information dependency between iterations

Each loop iteration has to depend from previous immediate iteration, otherwise iteration will be just a repetition of the previous iteration (memory used by iteration doesn't change) and hence, loop will run infinitely long. So iterations of whole loop body can't be executed in parallel, but parts of it can be. Iterations of level 1 controller have to have connections with body iterations; otherwise it can be run only one time.

For body with an indexed variable the result created on one iteration $n_0$ can be used on another iteration $n_1$ ($n_0<n_1$), not immediate next to $n_0$. To determine that the value of $a[K_0(i|n_0)]$ created on iteration $n_0$ used with indexed variable $a[K_1(j|n_1)]$ on iteration $n_1$ we have to solve an equation
$$K_0(i|n_0) = K_1(j|n_1)$$
for $n_0$ and $n_1$. The expression $i|n_0$ means value of i on iteration $n_0$.

For this equation we have to identify the system execution steps of the nested loops.

We'll use the following notations for nested loops and its elements. If $C_{i1}$ is a loop with body $B_{i1}$, then loops that belong to it will be denoted as $C_{i1,i2}$. In general $C_{i1,...,in}$ will denote loops belonging to the body of the loop $C_{i1,...,in-1}$. So depth of the nested loop is reflected in the index dimension of its name. The next diagram illustrates this notation:

Figure 4.1.

For simplicity we suppose that in a schema all instructions are different and each loop has a counter of iterations starting at 0 when loop instruction is initialized. Then for loop $C_{i1,...,in}$ a vector of counters $m=m_1,...,m_n$ of the inner loops $C_{i1,...,ip}$ for $p<n+1$ will be unique for the step of execution of any instruction q from its body and we will use the notation $q^m$.

### 4.1. Connection equation

Immediate connection between steps $q_1^{m1}$ and $q_2^{m2}$ (let $q_1^{m1}<q_2^{m2}$) takes a place when there is a simple or indexed variable value of it is created on the step $q1^{m1}$ and used on the step $q_2^{m2}$, or the result of $q1^{m1}$ is overwritten with $q_2^{m2}$ and it is the nearest step with such properties (there are no any operator $q_3$ and iteration $m_3$ ($m_3<m_2$) in between have a property like $q_2$).

For case of simple variable the nearest $m_2$ is the next iteration after vector $m_1$. In case of indexed variables $a[K_1(i)] \in Val(q_1)$ and $a[K_2(j)] \in Arg(q_2)$, immediate connection means that $K_1(i|m_1)=K_2(j|m_2)$ and there is no any instruction $q_3^{m3}$, where $q_1^{m1}<q_3^{m3}<q_2^{m2}$, with such a variable $a[K_3(p)] \in Val(q_3)$ that $K_3(p|m_3) = K_1(j|m_1)$ for $m_1<m_3<m_2$; $i|m_1$ means value for i calculated by some controller on iteration $m_1$.

So to detect information connections we have to have solution for following equation:
**$K_1(i|m_1)=K_2(j|m_2).$**

We will call it a **connection equation.**

It is a natural number equation which is a superposition of index expressions and functions of controllers on both sides. We can solve this problem for the system of linear equations but not for polynomials higher than 3 degrees.

It means that class of programs for solving system of linear equations, matrix operations and differential equations can be parallelized automatically [1,2]. But this way is a dead end. Even for a case when connection equation is a polynomial, solving algorithm does not exist: we have a Diofant's equation problem.

**Remark 4.1.** We can show that the equation is the only problem – if we have solution for connection equation, no more algorithmically unsolvable problems are left for parallelizing of free program schemas.
This is technical result and we are not going to show it here.

### 4.2. Avoiding the connection equation

Let modify programming language (and call it language of **programs with predecessors**) by changing only semantics of index variables: instead of asking connection equation we may ask its solution. Now a variable with index expression like $a[g(i)]$ may be used only in right side of assignment operator and it means that on current iteration must be used value for $a$ that was created on previous iteration $g(i)$ and hence $g(i)<i$. Other word, index expression reflects a point in iteration space, not a point in an array.

There is no needs to show an index expression in any left side, it always the current iteration (when this operator was executed).

**Example 4.1.** Let's have a 4-point deferential approximation:

$$f^k(i,j)=1/4(f^k(i-1, j)+f^k(i, j-1)+f^{k-1}( i+1, j)+f^{k-1}(i, j+1)).$$

The equivalent program schema with predecessors is next system of loops

```
for (k=1;k<N+1;k++)
    for (i=1;i<N+1;i++)
        for (j=1;j<N+1;j++)
            f = 1/4(f[k, i-1, j]+f[k, i, j-1]+f[k-1, i+1,j]+f[k-1, i, j+1]);
```

In reality last line has to be more complex because of start data. Start data or elements of input arrays might be introduced as negative number indexes. But for simplicity we are not showing calculation along the coordinate plains.

The natural way to execute a program with predecessors is to repeatedly run in parallel all iterations of loop's body fragments which have data. Obviously more efficient execution is to look out only iterations that just got data.

For example above controllers are simple and just increase indexes j, i, k. Their values do not depend on kernel (it a last line of code) and controller iterations can be executed before any kernel iterations. So we will get a list of 3D vectors (1,1,1), (1,1,2),…, (1,1,N),…, (1,2,1),…, (1.2,N), …, (1,N,N),…, (N,1,1),…, (N,N,N). For each of them we check if there is data for the arguments. In practice we have to check only points changed on a previous step.
At the beginning it is the point (1,1,1). Then data will be ready for 3 points (1,2,1), (1,1,2), (2,1,1).

Let P be a plain having these tree points, and **Norm** is its normal vector. Then at the next step data will be ready for iteration lying on next nearest plain with the same normal vector and including whole number coordinates. One can see that each next available for execution set of iterations is belonged to next parallel plain with integer points.

Each plain iterations on can be calculated in parallel. Really, for any integer point on the plain, all points used for this iteration are in an area strictly below this plain. It means that these points were already calculated and any point of plain can't be argument for another point of the same plain and hence points on one plain can be calculated in parallel.

So this 4-point approximation is highly parallel and for this 3D task parallel execution time is linear function of the dimension size.
Next figure 4.2 illustrates this case.

Figure 4.2.

Remarkable feature of this language is that it has a trivial algorithmically solvable problem of finding nearest iteration in which result of current iteration will be used. The index expression of right side variable is a solution of a connection equation. Still both languages are **algorithmically full**. It is easy to prove that there exists a fixed interpretation of functional and predicate symbols so that each algorithmic function can be calculated with a program from these classes.

There is no contradiction to Rise's theorem about insolvability of mass problems. The class of our programs is not an exact subclass – it is a full class.

For programs we have to consider two dimensions of fullness: algorithmically when any partial recursion function has a program for it calculation and **data fullness** when data selection functions are an algorithmically full class, any data structure can be represented.

Loops with simple controllers are very suited for parallelizing. They produced well organized data streams that can be efficiently implemented in parallel systems with separated memories. Questions are how useful such programs, how to organize parallel execution and how complex is it. The only comprehensive but expensive answer for that questions is a building of a real hardware and software system and applications of a wide enough area that executed in the system. But as it is shown in [3] there are many problems as well.

### 4.3. Cone of dependency and parallelism

Program with predecessors has a simple geometric interpretation. Let's consider hierarchy of loops with body B of the deepest loop. Then for each iteration $i_1,\ldots,i_n$ we may calculate all immediate predecessors which values are used on this step. Repeating this for each predecessor, for predecessors of predecessors and so on we will get a set of iteration which we call a **cone of dependency**. It is clear that to get result for the current iteration we need results of any iteration from this cone.

Figure 4.2 represents the cone for iteration (k,i,j) for the 4-point example above:

Figure 4.2.

Any dependency cones for any point of any plain of iterations executable in parallel do not include each others, so they are independent from each other. This plain is a tangential to the cone at point (k, i, j) and the cone lies behind one side of it.

The relationship between sets of independent and sets of iterations executable in parallel is not simple.

Next example shows that **even if any set of iterations are lying on any lines, parallel to some given one, are independent the parallel execution for such sets does not exist**.

**Example 4.2.**

```
For (z=1; z<N; z++)
{
    For (x=1; x<N; x++)
    {
        For (y=1; y<N; y++)
        {
            for (p=1; ((x+p<n)&(y+p<n)); p++)
            {
                m₀: if (x>y) then m1 else m₂
                m₁:v=f1(u[p,x+p], u[x+p,p],v) then m₂
                m₂:v=f2(u[p,y+p], u[y,y+p],v) then m₃
                m₃: u[x,y]=f3(u[x,y], u[x-1,y-1],v) then mₛ
            }
        }
```

```
            }
    }
```

The cone of dependency for these nested loops is shown in figure the 4.4.

    Here any set of iterations parallel to L (bold line) consists of independent iterations. But the two lines cannot be executed in parallel because dependency cone of one of them will contain points of another one. Thus any parallel to L line is a set of independent iterations, but only one line iterations can be executed in parallel.

    Figure 4.4.

    It can be proved that any forward loops can be transform to loops in separated form with the same technique as for ordinary program schemas. Number of its controllers and their connection topologies are an invariant for transformation. Therefore loops with different characteristics like these can't be equivalent.

    Important feature of the forward loops is its **data selection fullness**. The constructors may represent any functions. Hence any data structure might be represented and processed in parallel.

    **Conclusion.** We've shown that problem of loop iterations independency contains the problem of solving connection equations. The last problem might be avoided by changing semantic of index expressions. The getting class is algorithmically full and has full data selections.

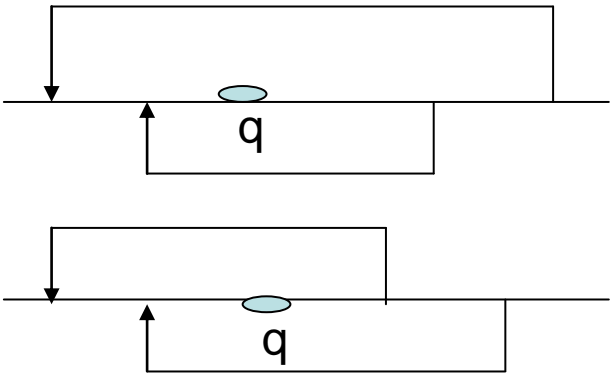
Figure 1.1. Structured and unstructured loops

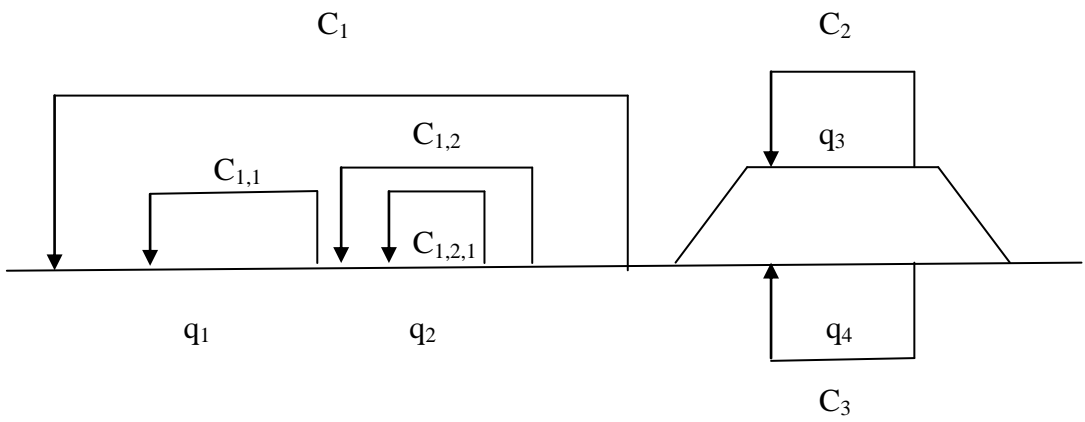

Figure 4.1. Loop naming

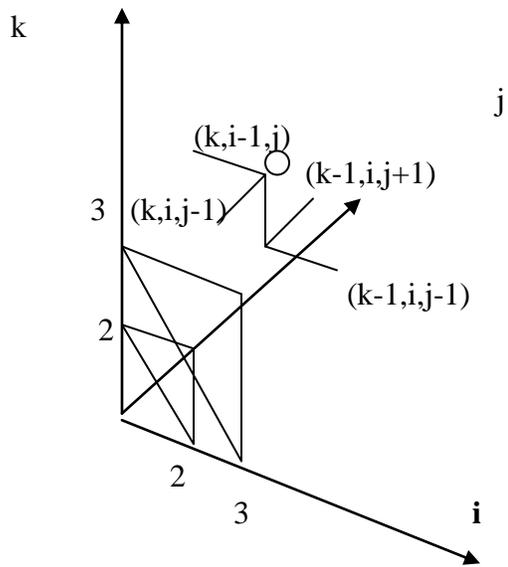

Figure 4.2. Hyper plains for parallel execution

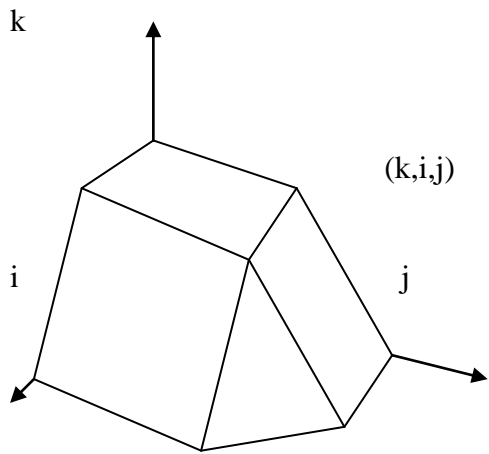

Figure 4.3. A cone for point (k,i,j)

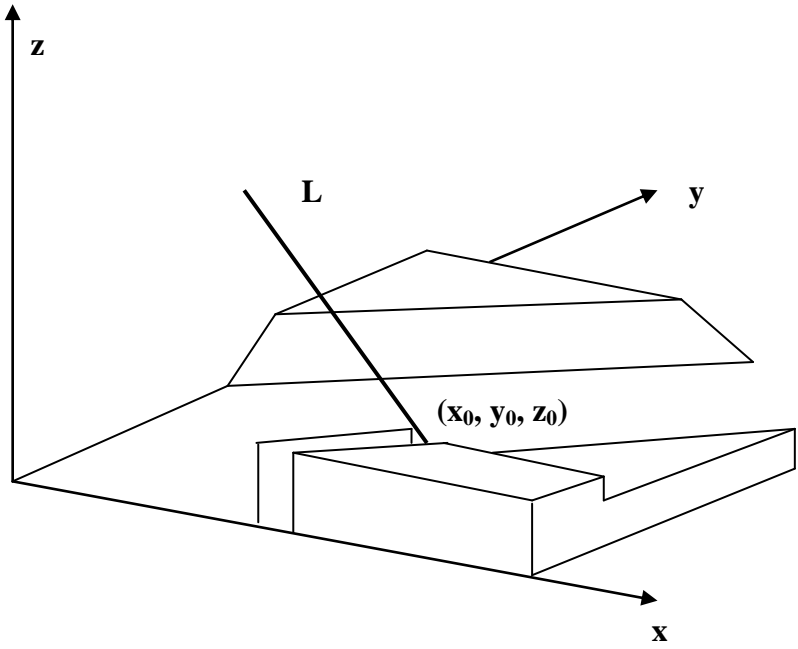

Figure 4.4. Cone of dependence for system from an example 2.4.2.